\documentclass[twocolumn]{aastex63}

\usepackage{amsmath}
\received{}
\revised{}
\accepted{}
\submitjournal{ApJL}

\shorttitle{Recent Star-formation in a Red Massive Lensed Quiescent Galaxy at z=1.88}
\shortauthors{M. Akhshik, K. Whitaker, J. Leja, et al.}

\graphicspath{{./}{figures/}}

\begin{document}

\title{Recent Star-formation in a Massive Slowly-Quenched Lensed Quiescent Galaxy at z=1.88}

\correspondingauthor{Mohammad Akhshik}
\email{mohammad.akhshik@uconn.edu}

\author[0000-0002-3240-7660]{Mohammad Akhshik}
\affil{Department of Physics, University of Connecticut,
Storrs, CT 06269, USA}

\author[0000-0001-7160-3632]{Katherine E. Whitaker}
\affil{Department of Astronomy, University of Massachusetts, Amherst, MA 01003, USA}
\affil{Cosmic Dawn Center (DAWN), Denmark}

\author[0000-0001-6755-1315]{Joel Leja}
\affil{Department of Physics, 104 Davey Lab, The Pennsylvania State University, University Park, PA 16802, USA}

\author[0000-0003-3266-2001]{Guillaume Mahler}
\affil{Department of Astronomy, University of Michigan, 1085 South University Ave, Ann Arbor, MI 48109, USA}

\author[0000-0002-7559-0864]{Keren Sharon}
\affil{Department of Astronomy, University of Michigan, 1085 South University Ave, Ann Arbor, MI 48109, USA}

\author[0000-0003-2680-005X]{Gabriel Brammer}
\affil{Cosmic Dawn Center (DAWN), Denmark}
\affil{Niels Bohr Institute, University of Copenhagen, Lyngbyvej 2, DK-2100 Copenhagen, Denmark}

\author[0000-0003-3631-7176]{Sune Toft}
\affil{Cosmic Dawn Center (DAWN), Denmark}
\affil{Niels Bohr Institute, University of Copenhagen, Lyngbyvej 2, DK-2100 Copenhagen, Denmark}

\author[0000-0001-5063-8254]{Rachel Bezanson}
\affil{Department of Physics and Astronomy, University of Pittsburgh, Pittsburgh, PA 15260, USA}

\author[0000-0003-2475-124X]{Allison Man}
\affil{Dunlap Institute for Astronomy and Astrophysics, University of Toronto, 50 St George Street, Toronto ON, M5S 3H4, Canada}

\author[0000-0002-7524-374X]{Erica J. Nelson}
\affil{Department of Astrophysical and Planetary Sciences, 391 UCB, University of Colorado, Boulder, CO 80309-0391, USA}

\author[0000-0003-4196-0617]{Camilla Pacifici}
\affil{Space Telescope Science Institute, 3700 San Martin Drive, Baltimore MD 21218, USA}

\author[0000-0002-3977-2724]{Sarah Wellons}
\affil{Center for Interdisciplinary Exploration and Research in Astrophysics (CIERA) and Department of Physics and Astronomy, Northwestern University, Evanston, IL 60208, USA}

\author[0000-0003-2919-7495]{Christina C. Williams}
\affil{NSF Astronomy and Astrophysics Postdoctoral Fellow}
\affil{Steward Observatory, University of Arizona, 933 North Cherry Avenue, Tucson, AZ 85721, USA}

\begin{abstract}

In this letter, we reconstruct the formation pathway of MRG-S0851, a massive, $\log M_*/M_\odot=11.02\pm0.04$, strongly lensed, red, galaxy at $z=1.883\pm0.001$. While the global photometry and spatially-resolved outskirts of MRG-S0851 imply an early-formation scenario with a slowly decreasing or constant star-formation history, a joint fit of 2D grism spectroscopy and photometry reveals a more complex scenario:  MRG-S0851 is likely to be experiencing a centrally-concentrated rejuvenation in the inner $\sim$1 kpc in the last $\sim$100 Myr of evolution.  We estimate $0.5\pm0.1\%$ of the total stellar mass is formed in this phase. Rejuvenation episodes are suggested to be infrequent for massive galaxies at $z\sim2$, but as our analyses indicate, more examples of complex star-formation histories may yet be hidden within existing data. By adding a FUV color criterion to the standard U-V/V-J diagnostic — thereby heightening our sensitivity to recent star formation — we show that we can select populations of galaxies with similar spectral energy distributions to that of MRG-S0851, but note that deep follow-up spectroscopic observations and/or spatially resolved analyses are necessary to robustly confirm the rejuvenation of these candidates. Using our criteria with MRG-S0851 as a prototype, we estimate that $\sim$1\% of massive quiescent galaxies at $1<z<2$ are potentially rejuvenating.

\end{abstract}

\keywords{}

\section{Introduction \label{sec:intro}} Galaxy colors are useful parameters to understand galaxy formation and evolution observationally, as they are relatively easy to measure for a large number of galaxies and a powerful tool to classify different galaxy types. Bimodality in both rest-frame color-color and color-magnitude planes is observed in the local universe \citep[e.g.,][]{strateva2001}, as well as out to at least $z\sim2$ \citep[e.g.,][]{williams2009,whitaker2011,whitaker2014,arnouts2013,mourtad2018}, with the main challenge in estimating robust rest-frame colors being the photometric redshift accuracy \citep{whitaker2010}.

Improved photometry has revealed trends within rest-frame color parameter space with various physical galaxy properties, such as age, specific star-formation rate (sSFR) and dust attenuation \citep[e.g.,][]{whitaker2017,fang2018,belli2019,leja2019,leja2019b}. Two common choices for studying these trends include the rest-frame U-V/V-J, NUV-r/r-J(K) diagrams (hereafter abbreviated UVJ and NUVrJ) color-color diagrams. A class of trajectories in these rest-frame color-color diagrams suggests that there exists a fast-quenching pathway with $\tau \sim 100$~Myr that includes compact quiescent galaxies evolving via dramatic events such as gas-rich mergers, as well as a slow-quenching pathway with $\tau\sim 1$~Gyr that generates larger quiescent galaxies \citep[e.g.,][]{mourtad2018,belli2019,carnall2019,rodriguez2019}. Rest-frame color-color diagrams therefore provide a valuable context toward understanding viable formation pathways of quiescent galaxies at $z\sim$1-2, one of the greatest challenges of observational extragalactic astronomy.

Motivated by the emerging picture in simulations connecting age and sSFR gradients to physical processes happening prior to quenching \citep[e.g.,][]{wellons2015,nelson2016, tacchella2016}, one can in principle spatially resolve rest-frame colors to understand the formation patterns within individual galaxies. However, analyzing spatially resolved properties of galaxies at $z\sim2$ is challenging due to spatial resolution mismatch across the wavelength ranges of interest. For instance, for a galaxy at $z\sim2$, both rest-frame J and K bands fall outside of the wavelength coverage of the high-resolution \emph{Hubble Space Telescope} (\emph{HST}) imaging. This gap can be covered by \emph{Spitzer} IRAC channels 1 and 2, albeit at significantly lower spatial resolution, with a few methods developed to address the resolution mismatch problem \citep[e.g.,][]{wuyts2012,akhshik2020}. Another technical difficulty for spatially resolving quiescent galaxies at $z\sim2$ stems from their relative compactness, as they are often barely resolved by even \emph{HST} \citep[e.g.,][]{vanderwel2014}. Strong gravitational lensing is a natural tool to study their stellar populations by boosting the signal-to-noise ratio and by extreme stretching of galaxies through the accompanied shear in some cases. 

In this letter, we present a spatially resolved stellar population analysis of MRG-S0851, a massive lensed red galaxy at $z=1.88$. The core methodology for fitting the stellar population of MRG-S0851 is presented in an accompanying paper \citep{akhshik2020}. Here, we discuss the reconstructed spatially-resolved star-formation histories (SFHs) and consider this target through the lens of rest-frame color-color diagrams, demonstrating their potential and limitations to interpret the seemingly complicated SFH of MRG-S0851. We first discuss the spatially-resolved SFHs from the joint spectrophotometric fitting in Section \ref{sec:spatially-resolved}. In Section \ref{sec:color-color}, we use the global photometry and the global rest-frame colors to compare the SED of MRG-S0851 to that of star-forming and quiescent galaxies selected using UVJ and a longer lever arm FUV-V/V-J (hereafter FUVVJ) color-color diagrams \citep{leja2019b}, from the 3D-HST galaxy survey, proposing a population that may have a similar formation pathway. Finally, in Section \ref{sec:discussion}, we interpret our results using the UVJ and FUVVJ color planes and combine all information to speculate about the evolution of MRG-S0851.

In this letter, we assume a standard simplified $\mathrm{\Lambda CDM}$ cosmology with $\Omega _{M}=0.3$, $\Omega _{\Lambda}=0.7$ and $H_0 = 70 \, \mathrm{km}\mathrm{s}^{-1}\mathrm{Mpc}^{-1}$ and the \citet{chabrier2003} initial mass function. All magnitudes are reported in the AB system.

\section{Spatially Resolved Star-Formation Histories \label{sec:spatially-resolved}}

The spatially-resolved star-formation history of MRG-S0851 is measured using the \texttt{requiem2d} code. We provide a brief summary in this section, and refer the reader to \citet{akhshik2020} for an in-depth discussion of our methodology.

\begin{figure}
    \centering
    \includegraphics[width=1.0\columnwidth]{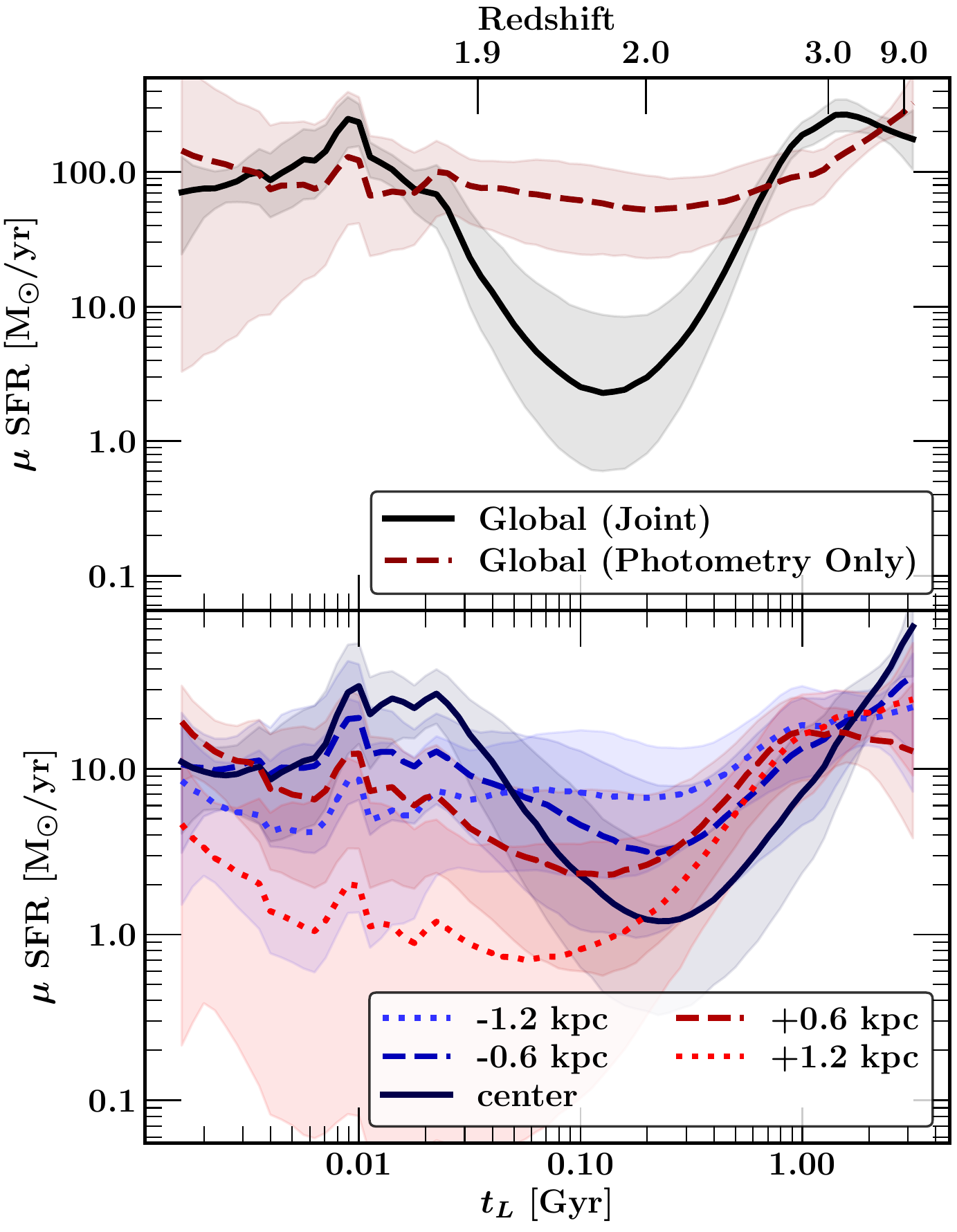}
    \caption{Reconstructed SFHs of MRG-S0851 as a function of lookback time, $t_L$, demonstrating how photometry-only global fit misses the enhancement of SFH in the last $\sim$100 Myr of evolution. The bottom panel shows the spatially-resolved model with global SSP prior, noting that the resolved SSP prior yields a consistent result within 1$\sigma$ for 5 central bins. The SFHs from the joint fit are generally decreasing from $z\sim9$ to $z\sim2$, but the central bins' and the global stellar populations show an increase in the SFR in the last $\sim$100 Myr of evolution. The SFRs are not corrected for gravitational lensing magnification $\mu$.}
    \label{fig:sfr-plot}
\end{figure}

\begin{figure}[!t]
    \centering
    \includegraphics[width=1.05\columnwidth]{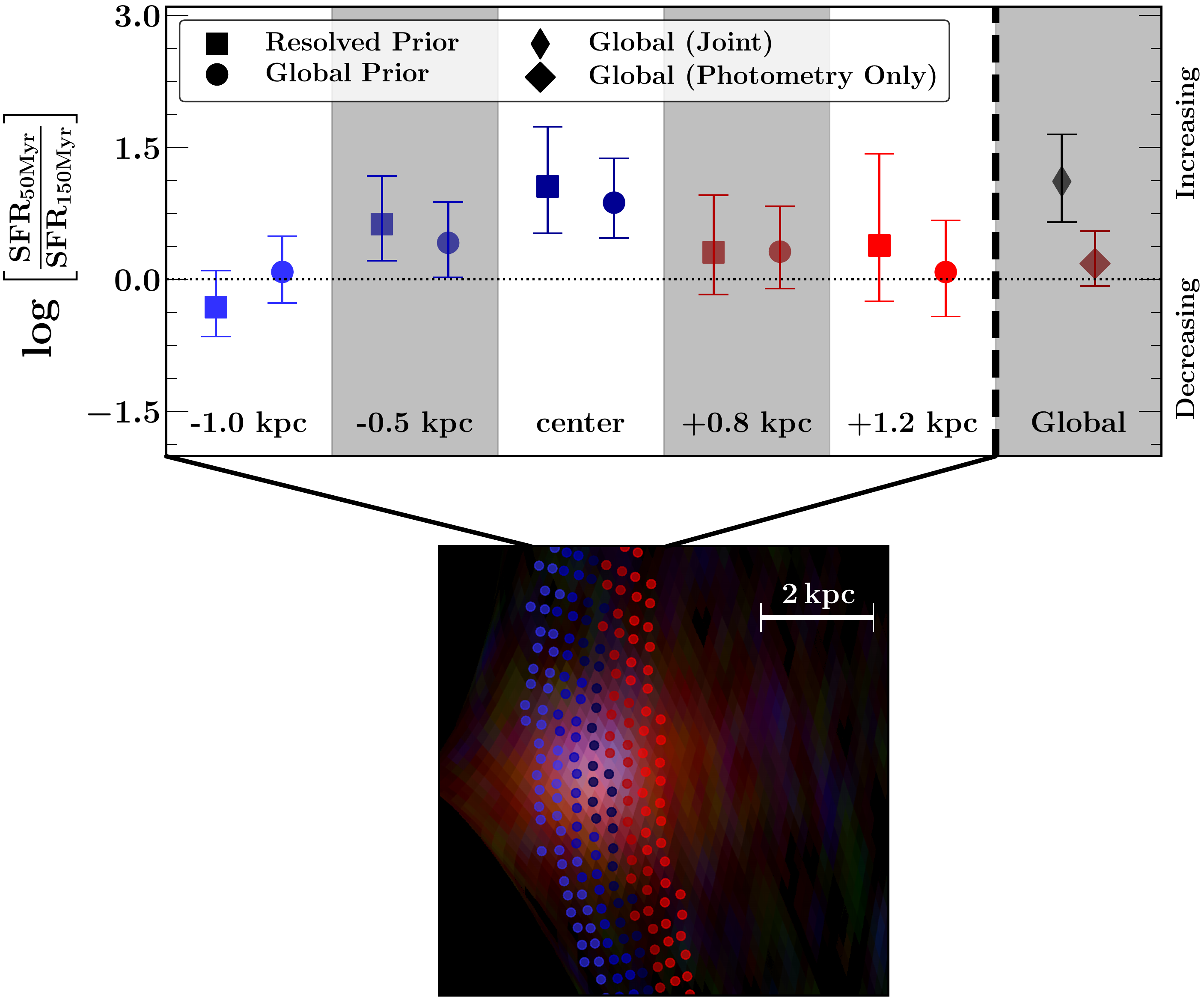}
    \caption{The ratio of SFR in lookback times of $\sim$50 Myr to $\sim$150 Myr, demonstrating the enhancement in SFR in the last $\sim$100 Myr of evolution in the global joint-fit and central bins. The bottom panel is the reconstructed color image of MRG-S0851, using the lensing and \texttt{GALFIT} model \citep{akhshik2020}. The location of pixels that outline each spatial bin are shown using the same color as the top panel.  The data points with different priors are offset spatially for clarity. The dashed horizontal line indicates no enhancement.
    }
    \label{fig:sfr-jump}
\end{figure}

The spectrophotometric data is jointly fit to constrain the SFH and age gradients using a linear combination of templates obtained by averaging simple stellar populations (SSPs). To model the dust/metallicity uncertainty, the posterior on dust and metallicity, estimated by fitting photometry data \citep[\texttt{Prospector-$\alpha$};][]{leja2017}, is divided into 12 regions. A set of templates is assigned to each region by averaging over 15 templates that are constructed by randomly drawing dust/metallicity from each region. The weight of each set is determined simultaneously with other free parameters. The SSPs have ages from 1 Myr to the age of the universe at $z=1.88$, increasing with a logarithmic steps of $\Delta \log t [\mathrm{Gyr}] = 0.05$, including nebular emission lines by assuming an ionization parameter of $\log U=-2.5$. A non-parametrized SFH, with a continuity prior is assumed \citep[e.g.,][]{leja2019,akhshik2020}.

We perform the joint-fit on both the global and spatially-resolved data. For the resolved analysis, we define 7 bins, probing $<$-1.4kpc to $>$+1.5kpc from the centroid along the semi-major axis, with average steps of $\sim$0.6~kpc in the source plane. The bins are not perfectly symmetrical because of the the non-uniform gravitational magnification. We test two ``SSP priors'' for the resolved analysis, corresponding to templates generated from the global and resolved \texttt{Prospector-$\alpha$} dust/metallicity posteriors, indicated with ``Global SSP Prior'' and ``Resolved SSP Prior''. The fits for the outermost two bins are not conclusive, as their grism spectra are contaminated by neighbouring sources.

The fit to the global photometric data suggests an almost constant SFH as a function of look-back time (Figure~\ref{fig:sfr-plot}, top panel). However, adding 2D grism spectroscopy reveals a more complex SFH for both the resolved and the global stellar populations. The SFHs for the 3 central bins (inner $\sim$1 kpc), as well as the global SFH from the joint-fit, show an enhancement of SFR in the last $\sim$100 Myr of evolution. Meanwhile, the spatial bins at +1.2~kpc and -1.0~kpc are consistent with a constant or slowly decaying SFH. Therefore, spatially resolved analyses and/or jointly fitting of 2D grism spectra changes our interpretation to instead be consistent with centrally-concentrated rejuvenation. In Figure \ref{fig:sfr-jump}, we calculate the ratio of $\mathrm{\log SFR_{50Myr}/SFR_{150Myr}}$ to further demonstrate the SFR enhancement in the last $\sim$100~Myr of evolution. The bottom panel shows the reconstructed color image in the source plane, where rejuvenation manifests as blue/purple pixels at the center.

Despite the recent rejuvenation, the global properties of MRG-S0851 are consistent with an early-formation scenario: MRG-S0851 has a stellar mass of $\log M_*/M_\odot=11.02\pm0.04$ after correcting for the gravitational magnification of $\mu = 5.7^{+0.4}_{-0.2}$, a global median mass-weighted age of $1.8_{-0.2}^{+0.3}$~Gyr (from \texttt{requiem2d}), and a compact size with a circularized effective radius of $r_c=1.7^{+0.3}_{-0.1}$~kpc \citep{akhshik2020}.  There is also no statistically significant radial variation in the median age of MRG-S0851 within the inner 3 kpc \citep{akhshik2020}.  Put together, the bulk of the stellar mass in MRG-S0851 likely formed very early in the universe at $z\gtrsim 2$, and quenching may have happened gradually without experiencing a late-stage central starburst event. We note that owing to the relatively flat SFH predicted from photometry alone, the added value of spectroscopy is not only to validate the rejuvenation hypothesis, but also to confirm that this galaxy quenched at some point between $z=1.9-3$. Both the resolved and global SFHs of MRG-S0851 show an extended period of early star-formation activity at $z\gtrsim 2$, consistent with a slow-quenching scenario \citep[e.g.,][]{carnall2019,belli2019}, with $t_{90}-t_{10}$ ($t_X$ indicates a time when X\% of total mass is formed) of $1.8^{+0.3}_{-0.2}$~Gyr for the unresolved stellar populations and $1.4^{+0.3}_{-0.4}$~Gyr (Resolved SSP Prior) and $1.5^{+0.3}_{-0.4}$~Gyr (Global SSP Prior) at the center. Rejuvenation occurred around $z\sim$1.9, with an estimated unlensed instantaneous $\mathrm{SFR_{<5 Myr}}=11.5_{-1.5}^{+1.7}~\mathrm{M_\odot/yr}$, calculated based on the best-fit nebular emission lines in SSPs at $t_L\lesssim15$~Myr driven by the $\mathrm{H\beta}$, $\mathrm{H\gamma}$, and $\mathrm{[O III]}$ emission lines within the grism data.

\begin{figure*}[!t]
\centering
\includegraphics[width=\textwidth]{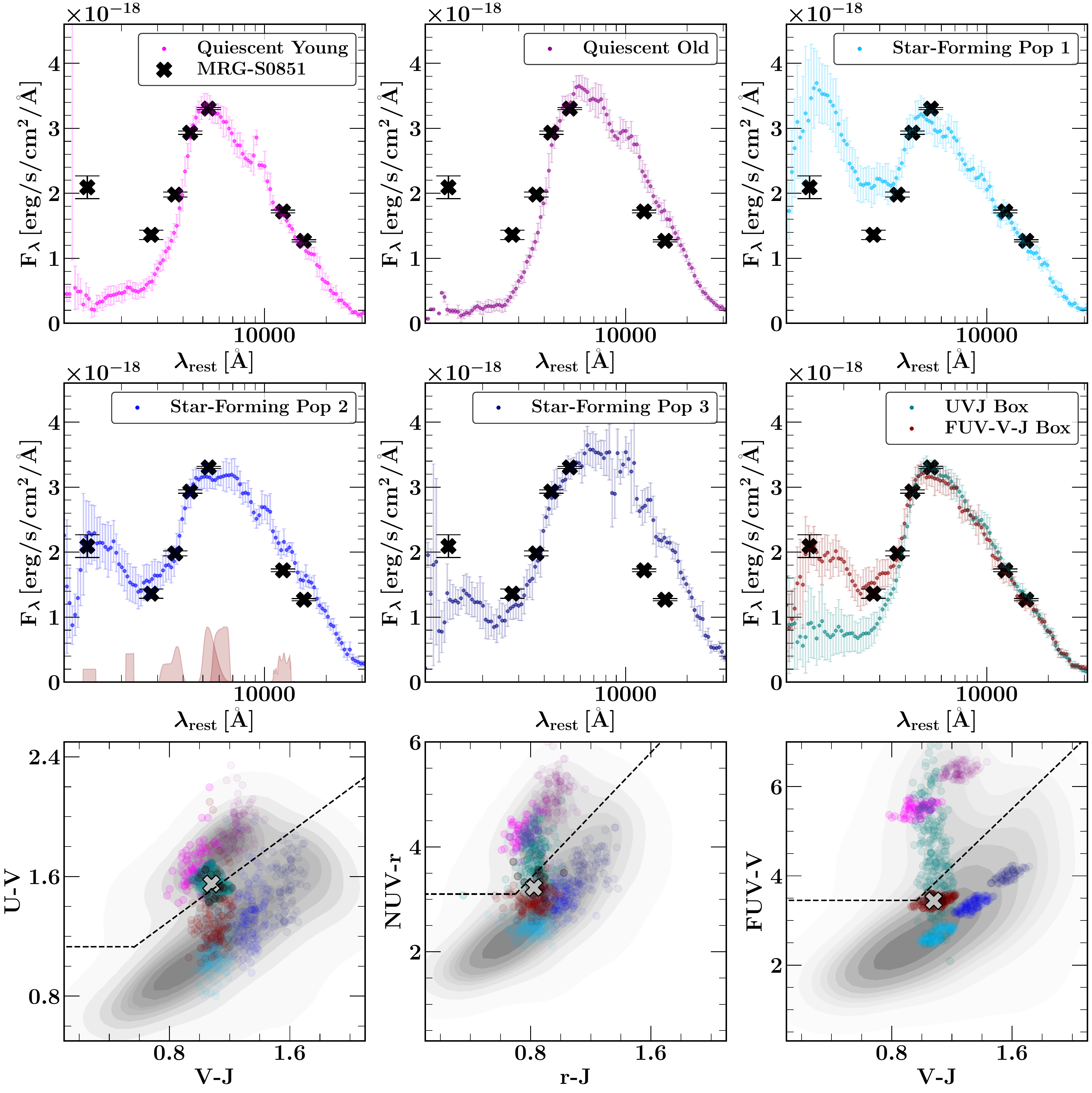}
\caption{The median and the median absolute deviation of the SEDs for seven populations of galaxies with $1.0<z<3.0$ and $\log M_* / M_\odot >10.0$ selected from 3D-HST, based on their location in the FUVVJ and UVJ rest-frame color-color diagrams. The full distribution of 3D-HST galaxies with $1.0<z<3.0$ and $\log M_* / M_\odot >10.0$ is shown with density maps in lower panels. The composite SEDs are shown with the same color of their corresponding population in the color-color diagrams. The photometric measurements of MRG-S0851 are shown with crosses. In the middle left panel, we show the throughput of different filters at the rest-frame, i.e. FUV ($\lambda_{\mathrm{pivot}}\sim1390\mathrm{\AA})$, NUV ($\lambda_{\mathrm{pivot}}\sim2199\mathrm{\AA})$, U ($\lambda_{\mathrm{pivot}}\sim3589\mathrm{\AA})$, V ($\lambda_{\mathrm{pivot}}\sim5479\mathrm{\AA})$, r ($\lambda_{\mathrm{pivot}}\sim6157\mathrm{\AA})$, and J ($\lambda_{\mathrm{pivot}}\sim12357\mathrm{\AA})$ from left to right (light red). The dashed lines separating star-forming and quiescent galaxies are drawn following \citet{mourtad2018,leja2019b}. \label{fig:candels}}
\end{figure*}

\section{Rest-frame Color-color Analysis \label{sec:color-color}}

As our reconstructed SFH shows, MRG-S0851 has two episodes of star-formation activity with different timescales. To link MRG-S0851 to a broader population of galaxies, we use rest-frame color-color diagrams, leveraging their strength to connect different time-scales of star-formation to rest-frame colors \citep[e.g.,][]{arnouts2013,leja2019b}. Out of three specific choices, namely UVJ, NUVrJ and FUVVJ, we concentrate on the rest-frame FUVVJ and UVJ color-color diagrams. In principle, one could instead use NUVrJ for our subsequent analysis and find similar results as the NUV filter is also sensitive to recent episodes of star-formation, albeit with larger scatter. However, FUV is correlated with sSFR over wider ranges of values \citep{leja2019b}. There is also a stronger synergy between FUV and U filters relative to NUV and U filters once we combine two color-color diagrams, potentially due to a wider wavelength range that the FUV and U bands provide in the rest-frame UV, providing more leverage for separating populations with different values of dust. 

We select six populations of galaxies from the rest-frame FUVVJ and one population from the rest-frame UVJ color-color diagram. Quiescent and star-forming populations of galaxies are identified from the 3D-HST galaxy survey \citep{brammer2012,momcheva2016,skelton2014} following \citet{leja2019b}. They are further divided into three populations of star-forming galaxies to demonstrate the effect of increasing dust on the composite SEDs, and two populations of quiescent galaxies corresponding to old and young quiescent galaxies \citep[e.g.,][]{whitaker2013,belli2019}. Finally, we select two populations by identifying galaxies with similar rest-frame colors to that of MRG-S0851, as measured by EAzY \citep{brammer2008}, in both rest-frame UVJ and FUVVJ color-color diagrams. 

We create composite SEDs for each population following \citet{kriek2011,forrest2018}, by normalizing all of the observed photometric data points of each 3D-HST galaxy with $1<z<3$ and $\log M_* / M_\odot >10$, as reported in the 3D-HST catalog, such that its average flux in the rest-frame $4500\mathrm{\AA}$ to $5500\mathrm{\AA}$ matches with the photometric data points of MRG-S0851 in the same wavelength range.  We then define a common rest-frame wavelength grid, calculating the median and median absolute deviation of all data points in each wavelength bin. Figure~\ref{fig:candels} demonstrates the composite SEDs as well as the locations of the populations on the rest-frame UVJ, FUVVJ, as well as NUVrJ color-color diagrams.

The composite SED shape for quiescent and star-forming galaxies is different than that of MRG-S0851 (Figure~\ref{fig:candels}, top and middle panels). The composite SED of galaxies directly overlapping with MRG-S0851 in the rest-frame UVJ fails to match the MRG-S0851 SED at the rest-frame UV wavelengths of $\lesssim 3000 \mathrm{\AA}$ (Figure~\ref{fig:candels}, middle right panel). We note that a $1\farcs5$ color aperture diameter in the image plane is used for MRG-S0851 whereas the equivalent color aperture diameter of $0\farcs7$ in 3D-HST would be $0\farcs7 \times \sqrt{\mu} \simeq 1\farcs7$, with a gravitational magnification of $\mu = 5.7^{+0.4}_{-0.2}$. By adopting a slightly larger aperture, this UV continuum flux decreases, getting slightly closer to the rest-frame UVJ selected composite SED. 

\begin{figure*}[!t]
\centering
\includegraphics[width=\textwidth]{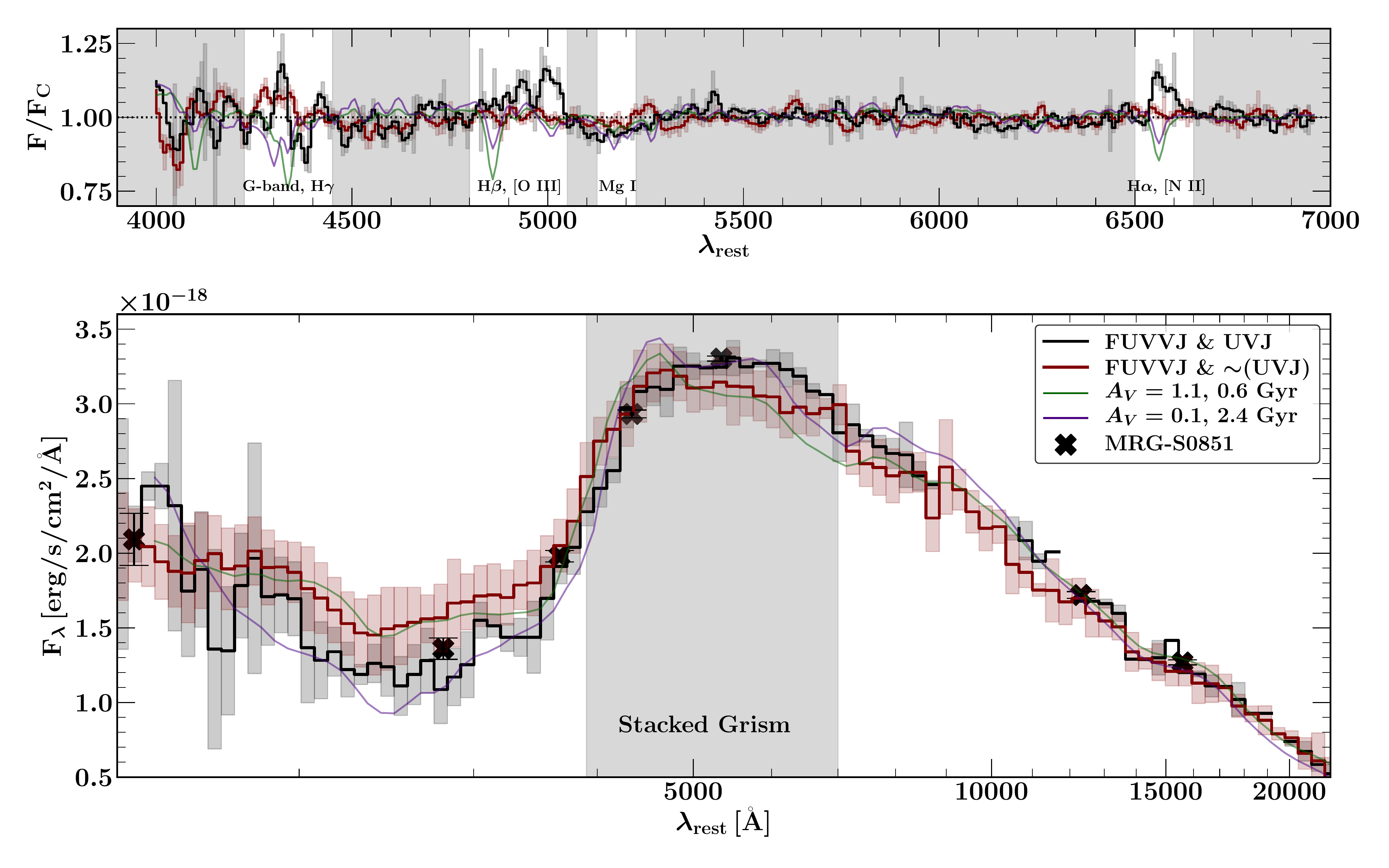}
\caption{The stacked grism spectra (top panel) and the composite SED (bottom panel) of FUVVJ and UVJ rest-frame color-color selected subpopulations from 3D-HST. The black curve in both panels shows a subpopulation with the same FUVVJ and UVJ rest-frame colors as MRG-S0851 (15 galaxies), and the red curve includes galaxies with the same FUVVJ but different UVJ colors relative to MRG-S0851 (85 galaxies). The purple and green lines are the best fits to these composite SEDs respectively. The population looks younger and dustier if we exclude the galaxies that have the same UVJ color as MRG-S0851. The weaker $\mathrm{H\alpha}$ emission line in the stacked G141 grism spectrum is consistent with a higher value of dust estimated for this population. \label{fig:stack}}
\end{figure*}

The SED of rest-frame FUVVJ color-color selected population is relatively well-matched with the MRG-S0851 SED (Figure~\ref{fig:candels}), notably around the rest-frame UV band $\mathrm{U_{F390W}}$. However, the population overlaps with both quiescent and star-forming galaxies in the UVJ color-color diagram (Figure~\ref{fig:candels}, bottom left panel). To investigate this scatter, we define two subpopulations using a synergy between the rest-frame UVJ and FUVVJ color-color diagrams: a subpopulation that matches with MRG-S0851 in both of these color-color diagrams and a subpopulation that matches MRG-S0851 in FUVVJ but not in UVJ. We calculate the composite SEDs of two subpopulations following the same steps as described above. For each of the galaxies in the subpopulations, we also continuum normalize the \emph{HST}/Wide Field Camera 3/G141 1D grism spectrum from 3D-HST, using a 5th degree polynomial while masking the emission/absorption lines \citep[e.g.,][]{whitaker2013}. The median grism spectra for each subpopulation is then calculated. We fit the composite SEDs following \citet{akhshik2020}, assuming solar metallicity and \citet{calzetti2000} dust attenuation law with $A_V=0.1\ldots2.1$. A $\chi^2$ analysis is performed to determine the age and $A_V$, with the best fits shown on both panels of Figure~\ref{fig:stack}.

As Figure~\ref{fig:stack} demonstrates, the FUVVJ rest-frame color-color selected population that has similar rest-frame colors to MRG-S0851 seems to include two different subpopulations, a young/dusty population that has a different UVJ color comparing to MRG-S0851 and an older population with significantly less dust that have similar UVJ colors to MRG-S0851. While the best fit SFHs of both sub-populations show signs of recent star-formation activity, their main period of star formation occurs at different times, leading to their different ages. We therefore propose that combining selections in both the FUVVJ and UVJ rest-frame color-color diagrams may yield rejuvenation candidates with a formation pathway consistent with MRG-S0851, a $\sim2$~Gyr old galaxy with rejuvenation in the last $\sim100$~Myr of evolution:
\begin{align}
        (\mathrm{U-V} &> 0.8(\mathrm{V-J}-1.08)+1.4) \, \mathrm{\&} \nonumber \\
        (\mathrm{U-V} &<0.8(\mathrm{V-J}-1.08)+1.7)\, \mathrm{\&}\nonumber  \\
        (\mathrm{U-V} &> -1.25(\mathrm{V-J}-1.08) + 1.4) \, \mathrm{\&} \nonumber \\ 
        (\mathrm{U-V} &< -1.25(\mathrm{V-J}-1.08)+1.7) \, \mathrm{\&} \nonumber \\
        (\mathrm{FUV-V} &< 2.8(\mathrm{V-J}-1.08)+3.74) \, \mathrm{\&} \nonumber \\
        (\mathrm{FUV-V}&>2.8(\mathrm{V-J} - 1.08)+3.14) \, \mathrm{\&} \nonumber \\ (\mathrm{FUV-V}&>-0.36(\mathrm{V-J}-1.08)+3.24) \, \mathrm{\&} \nonumber \\
        (\mathrm{FUV-V} &< -0.36(\mathrm{V-J} - 1.08)+3.64)
        \label{eq:sel}
\end{align}

We find a total number of 15 candidates that have clean grism spectra in the 3D-HST data release, also satisfying Equation \ref{eq:sel} at $1<z<2$. We, however, note that the rejuvenating candidates' selection rule in Equation \ref{eq:sel} is not complete, as it is obtained using the MRG-S0851 SED as a prototype. As the analysis in Section \ref{sec:spatially-resolved} also shows, deep spectroscopic observations and/or spatially resolved analyses are necessary to confirm that these candidate are rejuvenating. Simulating the SEDs of rejuvenating galaxies may shed light on the fidelity of the above selection rule as well. 
\section{Discussion \label{sec:discussion}}

\begin{figure*}[!t]
    \centering
    \includegraphics[width=1.0\textwidth]{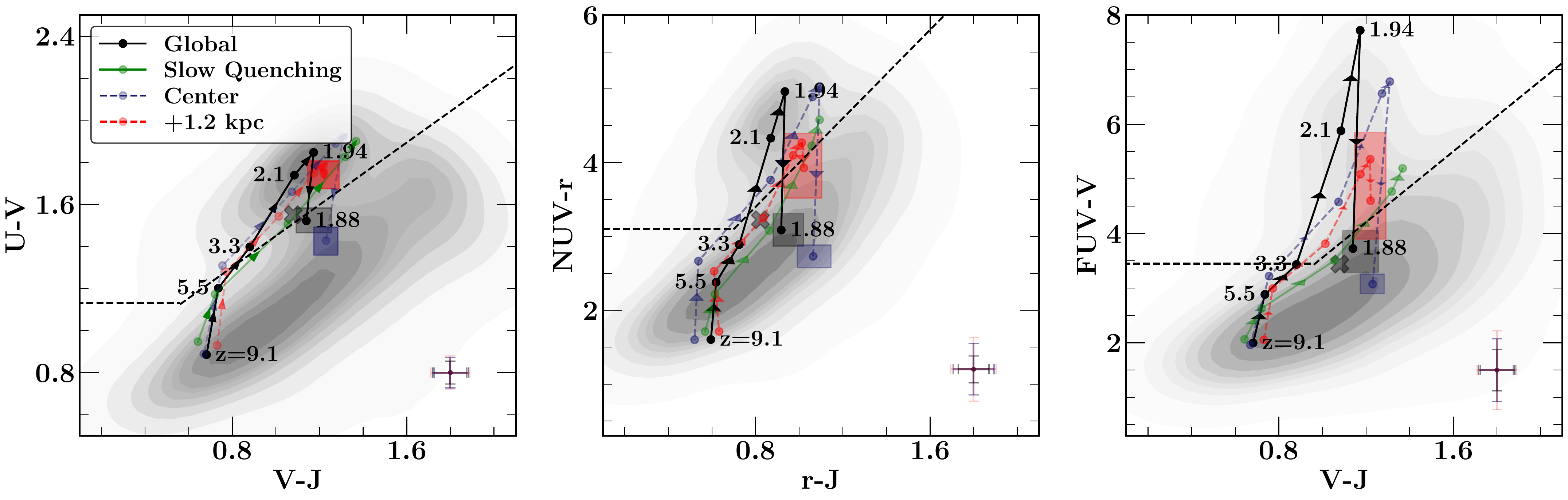}
    \caption{The reconstructed trajectories of MRG-S0851 for evolution in the rest-frame color-color diagrams, suggesting that the evolution of the global and spatially-resolved stellar populations at $z\gtrsim2.0$ and the outskirt bins at all redshifts are consistent with a slow-quenching formation pathway. The circles are snapshots at 6 redshifts, and the black cross is the global color as measured by EAzY. The color scheme for the trajectories is the same as Figure~\ref{fig:sfr-plot}, with black, dark blue and light red are used for global, central bin and the bin at +1.2kpc (adopting Global SSP prior). The boxes at the final redshift snapshot are 1$\sigma$ uncertainty, and the median uncertainties of different tracks are shown on bottom right. The green trajectory demonstrates a slow-quenching trajectory, constructed following \citet{belli2019}.}
    \label{fig:color-tracks}
\end{figure*}

To demonstrate the impact of a galaxy experiencing both an early formation and a late rejuvenation episode on the rest-frame color evolution, we use the posterior of the SFHs (Figure~\ref{fig:sfr-plot}), adopting the global SSP prior for the resolved analyses, to generate model spectra and calculate the colors in 6 redshift snapshots. We caution that no dust or metallicity evolution is assumed in calculating the tracks, with the color evolution caused solely by aging stellar populations. The result is shown in Figure~\ref{fig:color-tracks}, and to guide the eye, we construct a slow-quenching track following \citet{belli2019}, assuming a $\tau$-model for SFH with $\tau=650$~Myr. The trajectories of the spatially resolved bins and the global rest-frame color follow a slow-quenching pathway between $z\sim9$ to $z\sim2$, with MRG-S0851 entering the quiescent region in all rest-frame colors at $z\sim3$. The slow quenching scenario is also consistent with the duration of the star-formation activity, $t_{90}-t_{10}$, as discussed earlier. While recently quenched galaxies might have radial age gradients should they experience a central starburst event prior to quenching \citep{wellons2015}, we do not detect any statistically significant age gradients in inner $\sim$3 kpc of MRG-S0851 \citep{akhshik2020}. This suggests that it probably did not experience a rapid starburst event prior to the initial quenching at $z\sim3$, consistent with spectroscopic observations of other slowly-quenching galaxies at $z\sim2$ \citep{belli2019}.

The evolution of MRG-S0851 after $z\sim2$ seems to be different than a simple extension of the slow-quenching episode, as the central bin and global stellar population both shift toward bluer rest-frame optical colors between $z=1.94$ to $z=1.88$, with the $z=1.88$ snapshot overlapping with the star-forming population. However, the trajectories of the outer bins at +1.2~kpc and -1.0~kpc do not show this significant shift, ending in the quiescent region, given their uncertainty. Therefore, the reconstructed SFHs (Figure~\ref{fig:sfr-jump}) and the derived evolution in color-color diagrams support the idea that the most recent rejuvenation phase of MRG-S0851 is centrally concentrated. 

The current consensus is that minor mergers or the accretion of cold gas could trigger rejuvenation in otherwise quiescent galaxies \citep[e.g.,][]{yi2005,belli2017}. Given the fact that the rejuvenation is centrally concentrated in MRG-S0851, it might have been preceded by a gas-rich merger event or other dissipative mechanisms that funneled gas to the center. Rejuvenation is spatially resolved for local galaxies both in the outskirts, associating it with the infall of new gas or consumption of existing gas in the arms of spirals \citep[e.g.,][]{gonzalezdelgado2017}, or in the center because of the role of bars or other tidal interactions directing gas into the center \citep[e.g.,][]{chown2019}. While a minor merger may cause rejuvenation, young stars in a newly accreted object can also masquerade as rejuvenation. However, we find no indications of tidal disturbances down to the $25.1 \mathrm{mag/arcsec^2}$ surface brightness limit of our $\mathrm{H_{F160W}}$ observation, which should be sufficiently deep to detect moderate tidal features \citep{tal2009}. This suggests that MRG-S0851 is experiencing in-situ star formation rather than a very recent ($<$100~Myr) merger. 

An alternative explanation for the formation pathway of MRG-S0851 is an extended period of star-formation from $z\sim9-1.88$, similar to the SFH that we obtained from the photometry-only fit (Figure~\ref{fig:sfr-plot}). For example, \citet{thikler2010} show that the outermost part of NGC404 experienced recent star-formation activity.  The subsequent analysis of the chemical abundances of the ionized gas by \citet{bresolin2013} indicate similar metal enrichment to stars and therefore favors declining star-formation leftover from the primary star-formation episode. An analysis of the chemical abundance gradient in MRG-S0851 would therefore be a viable approach to ruling out the extended episode of star-formation, possible with the \emph{James Webb Space Telescope}. While our reconstructed resolved SFHs favor a rising SFH between $z\sim2$ to $z=1.88$ (Figure \ref{fig:sfr-jump}), this alternative scenario cannot be completely ruled out. Given the detection of nebular emission lines of [O III], $H\beta$ and $H\gamma$ in MRG-S0851 \citep{akhshik2020}, we constrain the gas-phase metallicity of $\log Z_{\mathrm{gas}}/Z_\odot=-0.45^{+0.04}_{-0.05}$, whereas the global stellar metallicity is constrained to be $\log Z_{\mathrm{star}}/Z_\odot=-0.09\pm0.04$ \citep{akhshik2020}. The gas-phase metallicity is $\sim$0.35 dex lower than the stellar metallicity, which could suggest that an accretion of lower-metallicity gas is a viable scenario. This result is consistent with the observed lower gas-phase metallicity in some local star-forming elliptical galaxies, suggested to be consistent with the accretion of gas from an external source \citep{davis2019}. We finally note that a UV upturn between Lyman-break and rest-frame 2500$\mathrm{\AA}$ is observed in low redshift early-type galaxies, and a potential scenario is the old horizontal branch stars becoming UV-bright  \citep[e.g.,][]{yi2004}. However, this scenario is unlikely for MRG-S0851, as the stellar population generating the UV upturn in a typical red sequence galaxy seems to be developed at $z\lesssim 0.7$ \citep{ali2018}, and hence they should be significantly older than the stellar populations of MRG-S0851.

Results from the Illustris simulation support a picture where rejuvenation phases last $\sim$1 Gyr, equivalent to the median time-scale of galaxies exiting and re-entering the red population in the color-stellar mass plane \citep{nelson2018}. As we have only had the opportunity to observe the enhancement for the last $\sim$100~Myr, MRG-S0851 may continue this rejuvenation phase for a significant period of time into the future.  We estimate that $0.5\pm0.1$\% of the total stellar-mass and $0.6^{+0.2}_{-0.1}$\% (Resolved Prior) and $0.7^{+0.2}_{-0.1}$\% (Global Prior) of the stellar-mass in the inner 3 kpc is formed in the last 100 Myr of evolution. With an average unlensed $\mathrm{SFR_{100Myr}}$ of $5 \pm 1 M_\odot/\mathrm{yr}$ in the last 100 Myr after correcting for gravitational magnification, we speculate that $\sim5\%$ of the total stellar-mass would be formed in a rejuvenation period should it last $\sim$1~Gyr with a steady SFR. This estimate is broadly consistent with that of $\leq 10$\% from \citet{chauke2019} for the fraction of stellar-mass formed during rejuvenation phases for a sample of quiescent galaxies selected from the LEGA-C survey \citep{vanderwel2016}.

Semi-analytic models predict that $\sim 30 \%$ of the quiescent galaxies evolving from $z=3$ to $z=0$ should experience a transient phase of rejuvenation \citep{pandya2017}. On the other hand, first results from the IllustrisTNG simulations predict that only 10\% of massive ($\log M_*/M_\odot > 11.0$), quiescent galaxies  will rejuvenate once and $\sim$1\% more than once \citep{nelson2018}. \citet{behroozi2019} use the \texttt{UNIVERSE\_MACHINE} model, constructed to determine galaxies' SFRs from their host halos, to show that the majority of galaxies at $z\sim 0$ with stellar masses of $\log M_*/M_\odot \sim 11$ have been rejuvenated once, with this fraction dropping sharply for lower and higher mass galaxies, as well as at higher redshifts.  Rejuvenation events like that observed in MRG-S0851 might be rare, and our study suggests that detecting the relevant observational signatures requires deep spectroscopy and/or high resolution spatially-resolved studies of $z\sim2$ galaxies. Specifically, the rejuvenating region of MRG-S0851 without gravitational lensing would only extend $\sim 0\farcs12$, which is smaller than the full-width at half-maximum of the $\mathrm{H_{F160W}}$ PSF. It would therefore be extremely difficult to detect for unlensed galaxies at $z\sim2$. Using the MRG-S0851 SED as a prototype to search FUVVJ and UVJ rest-frame color-color diagrams, we propose criteria to identify potentially rejuvenated candidates, estimating the ratio of the these candidates to the total number of quiescent galaxies to be $\sim$1\% with no significant variation at $1<z<2$ and $\log M_*/M_\odot>10$. This number is likely a lower limit, consistent with theoretical results as discussed above. We identify 15 potential early-formed rejuvenating candidates in the 3D-HST catalogs, and follow-up observations using deep grism spectroscopy and/or spatially resolved studies will confirm the fidelity of candidates identified using these criteria.

\section*{acknowledgements}
We would like to thank the anonymous referee for sharing valuable comments and questions that helped us improve the draft significantly. M.A. gratefully acknowledges support by NASA under award No 80NSSC19K1418. This work is based on observations made with the NASA/ESA Hubble Space Telescope, HST-GO-14622, obtained at the Space Telescope Science Institute, which is operated by the Association of Universities for Research in Astronomy, Inc., under NASA contract NAS 5-26555. K.W. wishes to acknowledge funding from the Alfred P. Sloan Foundation. S.T. and G.B. acknowledge support from the ERC Consolidator Grant funding scheme (project ConTExt, grant number No. 648179). The Cosmic Dawn Center is funded by the Danish National Research Foundation under grant No. 140.  C.C.W acknowledges support from the National Science Foundation Astronomy and Astrophysics Fellowship grant AST-1701546. The Dunlap Institute is funded through an endowment established by the David Dunlap family and the University of Toronto.

\bibliography{sample63}

\end{document}